\newif\iffigs\figsfalse

\documentstyle[12pt]{article}
\iffigs
  \input epsf
\else
\fi

\begin{document}


\def\a{\alpha}
\def\ad{$~\alpha~$}
\def\b{\beta}
\def\bd{$~\beta~$}
\def\mx{{\rm max}}
\def\mc{{\rm c}}
\def\Sr{$~\tilde{S}^{\rm red}~$}
\def\Srz{\tilde{S}^{{\rm red},2}}
\def\rum{r_{\mu_{\rm min}}}

\renewcommand{\thefootnote}{\fnsymbol{footnote}}
\renewcommand{\baselinestretch}{1.3}


\thispagestyle{empty}

\hbox to \hsize{%
  \vtop{\hbox{accepted for publication in }\hbox{\sl Phys. Lett. B}} \hfill
  \vtop{ \hbox{gr-qc/9406018}
\hbox{MPI-PhT/94-31}\hbox{June 1994}}}

\vspace*{1cm}

\bigskip\bigskip\begin{center}
{\bf \Huge{On the Stability of Gravitating Nonabelian
Monopoles}}
\end{center}  \vskip 1.0truecm
\centerline{\bf Helia Hollmann\footnote{e-mail:
hollmann@iws186.mppmu.mpg.de}}
\vskip5mm
\centerline{Max-Planck-Institut f\"{u}r
 Physik, Werner-Heisenberg-Institut}
\centerline{F\"ohringer Ring 6, 80805 Munich, Germany}
\vskip 2cm
\bigskip \nopagebreak \begin{abstract}
\noindent
The behaviour of magnetic monopole solutions
of the Einstein-Yang-Mills-Higgs equations subject to
linear spherically symmetric perturbations is studied.
Using Jacobi's criterion some of the monopoles are shown
to be unstable. Furthermore the numerical results and
analytical considerations indicate the existence of a
set of stable solutions.

\end{abstract}

\newpage\setcounter{page}1

\section{Introduction}

Stimulated by `t~Hooft's \cite{'tH74} and Polyakov's
\cite{Pol74} research van Nieuwenhuizen et al. \cite{NieWilPer76}
studied the influence of gravity
on magnetic monopoles, but their results do neither guarantee
the existence nor the stability
of the solutions as shown in \cite{BreForMai92},
\cite{Leeetal92}, \cite{Ort92}.
Later Breitenlohner et al. \cite{BreForMai92} and others
(Lee et al. \cite{Leeetal92}, Ortiz \cite{Ort92})
investigated `t~Hooft's and Polya\-kov's monopoles
(\cite{'tH74}, \cite{Pol74}) in curved space using
numerical methods.
Introducing the dimensionless parameters $~\a = M_W/g
M_{\rm Pl}~$ (with the Planck mass $~M_{\rm Pl} =
1/\sqrt{G}$) and  $~\b = M_H/M_W$, where  $~M_W~$ and
$~M_H~$ denote the mass of the corresponding Yang-Mills
field and Higgs field respectively, Breitenlohner et al.
\cite{BreForMai92} presented the following results:
For $~0 \le \b < \infty~$ the monopole mass varies only
over a finite domain. Choosing the parameter $~\b = 0~$
solutions seem to exist for $~0 \le \a \le \a_\mx$. For
$~\a_\mc \approx 1.386 \le \a \le 1.403 \approx \a_\mx~$
there are two solutions with different masses.  Considering
the mass as a function of $~\a$, one finds that it attains
a maximum $~M_\mx$ for  $~\a = \a_\mx$, and that it decreases
towards $~M_{\rm Pl}/g~$ asymptotically as $~\a$ tends
to $~\a_\mc$.  This situation is shown in {\small \bf Figure~1}.
Each point represents a solution calculated numerically.
The right plot of {\small \bf Figure~1} is a magnification of
the left plot near the point $~B$. The series of points
marked by $~BC~$ will be denoted by ``upper branch'',
the part $~AB~$ by ``lower branch''.

In this paper stability of the nonabelian gravitating magnetic
monopoles is investigated using the technique of
spherically symmetric perturbations. Stability considerations
of similar nature have been conducted by Zhou and Straumann
\cite{ZhoStr90}.
The stability problem requires
the numerical behaviour of the monopole solutions to
be known. The construction of the monopole solutions
in question (\cite{BreForMai92}) is briefly reviewed.
In order to investigate their stability the monopoles
are subjected to small perturbations in time,
which have a Fourier expansion due to its linearity.
Eliminating the gravitational degrees of freedom,
the remaining terms in the action are written
as a quadratic form. Then the application of Jacobi's
criterion (\cite{GelFom63}, \cite{Baa92}) suggests
instability on the branch $~BC~$ and stability under
spherically symmetric perturbations on the lower
branch. Note however that this does not imply
stability in general.  In the final section the
results are summarized and discussed.

\section{Solutions of the EYMH Equations}

The 't Hooft Polyakov monopole in curved space is a
static finite energy solution of an SO(3) gauge theory
with a triplet of scalar fields. Using a coordinate free
notation the action reads
\begin{equation}
\label{wifu}
S ~=~  \int - \frac{1}{16 \pi G} {*R} - \frac{1}{4 \pi}
       \left( - \frac{1}{4 g^2} Tr(F \wedge *F)
      + \frac{1}{2} Tr(D\Phi \wedge *D\Phi) \right) + U(\Phi),
\end{equation}

where $~R$, $F~$ and $~\Phi~$ denote the curvature
scalar, the Yang-Mills field and the scalar field
respectively. $G$ and $g$ are the gravitational
constant and the gauge coupling. $*$ is the Hodge
star operator and $~U(\Phi)~$ is a potential term
which will be specified later.  The gravitational
field is described by a static spherically symmetric
metric tensor given by the line element $~ds^2~$ in
Schwarzschild coordinates:
\[
ds^2 = A^2(r)\mu(r)~ dt^2 - \frac{1}{\mu(r)}~ dr^2
      - r^2 ( d\theta^2 + \sin^2\theta ~d\phi^2).
\]
The minimal ansatz for the Yang-Mills potential $~A~$ which
is spherically symmetric is denoted in polar coordinates
and abelian gauge as follows (\cite{WuYan69},\cite{ForMan80},
\cite{BreForMai93}):
\[
A~=~\left( W(r)~T_2 + \cot\theta~T_3 \right)
               \sin\theta~d\phi + W(r)~T_1~d\theta.
\]
The $~T_i$, $i = 1,2,3$, are the generators of the
fundamental representation of SO(3). Using the potential
$~A~$ the field $~F~$ is expressed by
\[F ~=~ dA + \frac{1}{2}~ [A,A].\]
The Higgs field is chosen to be
\[
\Phi~=~H(r) \left( \cos\phi \sin\theta~T_1
+ \sin\phi \sin\theta~T_2 + \cos\theta~T_3\right).
\]

In order to reduce the four dimensional action to a
one dimensional integral the curvature scalar, the
Yang-Mills, and the Higgs terms are substituted into
the action (\ref{wifu}):
\begin{equation}
\label{rerewifu}
S^{red} = - \int_0^{\infty} dr~A
\left(\frac{1}{2\a^2}(\mu + r\mu' - 1) - \mu V_1
- V_2\right),
\end{equation}

with
\begin{eqnarray}
V_1 & = & {W'}^2 + \frac{r^2}{2}{H'}^2,  \nonumber \\
V_2 & = & \frac{(W^2 - 1)^2}{2r^2} + W^2H^2 +
          \frac{1}{8}\b^2 r^2(H^2 - 1)^2.\nonumber
\end{eqnarray}

The term $~\frac{1}{8}\b^2 r^2(H^2 - 1)^2~$ represents
the potential $~U(\Phi)$.  The equations of motion
(EYMH equations) deduced from the action (\ref{rerewifu})
contain the parameters \ad and \bd as well. Their values
determine the stability properties of the monopoles.

The ``background solution'' $~{\bf F}(r)$
corresponding to global regular monopoles has
been obtained solving the equations of motion
with a multiple shooting method \cite{StoBul90}.

\section{Stability Analysis}

The static background solution $~{\bf F}(r)~$ is subjected
to a spherically symmetric perturbation $~{\bf f}(r) \mbox{e}^
{i \omega t}$, i.e. the perturbed and the
background solutions are related as follows:
\begin{equation}
\label{ssst}
{\bf \tilde{F}}(r,t) ~=~ {\bf F}(r)
{}~+~ {\bf f}(r) \mbox{e}^{i \omega t},
\end{equation}

where $~{\bf F}(r)~$ is the vector of the functions
$~H(r),~W(r),~Y(r)~$ and $~M(r)~$ and
$~{\bf f}(r)~$ denotes the vector $~(h,~v,~y,~m)~$ of
the radial parts of the perturbations.
It is convenient to introduce $~M(r)~$ and $~Y(r)$:
\[ \mu(r) ~=~ 1 - \frac{2 M(r)}{r}, \qquad \mbox{and}
\qquad A(r) ~=~ \mbox{e}^{Y(r)}.
\]

Instability in this context means the growth of the
exponential term in time. The analysis is thus reduced
to the question whether admissible perturbations, i.e.
perturbations fulfilling the boundary conditions,
with an imaginary frequency exist. Substituting
$~{\bf \tilde{F}}(r,t)~$
into the action (\ref{wifu}) one obtains a two dimensional
reduced action differing from (\ref{rerewifu}) only
by the time-dependent additional term
\[V_0 ~=~ \frac{1}{A \mu} \left( \dot{W}^2
      + \frac{r^2}{2}\dot{H}^{2} \right).
\]

The part of the reduced action \Sr quadratic in the
fields and their derivatives turns out to be
\begin{eqnarray}
\Srz \hspace{-0.3cm} &=& \hspace{-0.3cm}
    \int_0^{\infty} dr ~\mbox{e}^Y
     \left\{ y^2~ \left(\frac{M'}{\a^2}
    - \mu V_1 - V_2\right)  - 2y\mu~(2W'v' + r^2H'h')
    \right. \nonumber \\
&&\hspace{-1.7cm} + 2y\left(\frac{m'}{\a^2}
    \hspace{-1pt} +\frac{2 m}{r}V_1
   \hspace{-1pt} -\frac{2 W(W^2-1)}{r^2}v \hspace{-1pt}
   - 2WH^2v \hspace{-1pt} - 2W^2Hh
   \hspace{-1pt} -\frac{\b^2}{2} r^2 H(H^2-1)h \right)
    \nonumber \\
&&\hspace{-1.0cm} +~ \frac{4 m}{r} ( 2 W'v' + r^2 H'h')
    - \mu ~ (2v'^2 + r^2 h'^2) - \frac{2(3W^2-1)}{r^2} v^2
    -~ 2 H^2 v^2 \nonumber \\
&&\hspace{-1.0cm} \left. -~ 8 WH vh -2 W^2h^2
    - \frac{\b^2}{2} r^2(3H^2- 1)h^2
  -~ \omega^2 \frac{\mbox{e}^{-2Y}}{\mu} (2v^2 + r^2h^2)
    \right\}.\nonumber
\end{eqnarray}

Taking into account the boundary conditions at $~r = 0$,
the variation with respect to $~y~$ yields
\begin{equation}
\label{dgls}
\frac{m'}{\a^2} + \frac{m}{\a^2} Y' + D_{W'}L~v' + D_WL~v
+ D_{H'}L~h' + D_{H}L~h = 0,
\end{equation}

where $L~$ is the integrand of \Sr and $~D_{W'},~D_W,~D_{H'}$,
and $D_H~$ denote the derivatives with respect to $~W',~W,~H'$
and $~H$.
(\ref{dgls}) is solved by
\[m ~=~ \a^2~(2\mu W^{'} v + r^2 \mu H^{'} h).
\]

This result is used to eliminate $~m~$
and $~y~$ in $~\Srz$. Partial integration and introduction of
$~\hat{h} = r h / \sqrt{2}~$ and the variable $~\sigma~$
by $~dr = A\mu~d\sigma$ leads to
the normal form of a 2-channel Schr\"odinger problem $~\Srz$:
\begin{equation}
\label{2-K}
\Srz = \int_0^{\infty} d\sigma~
\left\{ D\Psi^{\dag} D\Psi ~+~ \Psi^{\dag} Q(r) \Psi
  ~-~ \omega^2 \Psi^{\dag} \Psi \right\}.
\end{equation}

Here
\begin{eqnarray}
\Psi^{\dag} &=& (\hat{h},~v), \quad
D\Psi^{\dag} ~=~ (\frac{d\hat{h}}{d\sigma},~\frac{dv}{d\sigma})
\nonumber \\
Q(r) &=& A \mu~ \left( \begin{array}{cc}
     Q_{11}(r) & Q_{12}(r) \\
     Q_{21}(r) & Q_{22}(r)
                  \end{array} \right) \nonumber \\
& & \nonumber \\
Q_{11}(r) &=& \frac{2W^2}{r^2} + \frac{\b^2}{2}(3H^2-1)
       + \frac{2\a^2}{r^2}\frac{1}{A} (A\mu r^3H'^2)'
        + \frac{1}{A} \frac{(A\mu)'}{r}
\nonumber \\
Q_{12}(r) &=&  \frac{2 \sqrt{2}}{r}
     \left(HW + \a^2\frac{1}{A}(A\mu r W'H')'\right)
{}~=~ Q_{21}(r)
\nonumber \\
Q_{22}(r) &=& H^2+\frac{3W^2-1}{r^2}
    + 4\a^2 \frac{1}{A} (\frac{A\mu W'^2}{r})'.
\nonumber
\end{eqnarray}

The background solution is unstable, iff the
Schr\"odinger problem admits a bound state,
that is a solution with $~\omega^2 < 0$.
The standard technique for a 2-channel Schr\"odinger
problem is to calculate the
equations of motion from (\ref{2-K}), which then define
an eigenvalue problem with the eigenvalue $~\omega^2$.
Instead of analyzing the spectrum of this operator the
positive definiteness of (\ref{2-K}) without the last
term is investigated
applying Jacobi's criterion \cite{GelFom63},
\cite{Baa92}. Although in \cite{GelFom63} it is formulated
for compact intervals only the proof
applies in this case to an infinite interval of integration, too,
in view of the asymptotic behaviour of $~\Psi$. The discussion of
Jacobi's criterion in \cite{GelFom63} also provides a normalizable
perturbation with negative energy.

This criterion states that the monopole solutions are
unstable iff
\begin{equation}
\label{det}
\det V(r) ~=~
\left|
\begin{array}{cc}
\hat{h}^{(1)}(r) & \hat{h}^{(2)}(r) \\
v^{(1)}(r) & v^{(2)}(r)
\end{array}
\right| ~=~ 0 \qquad \mbox{for} \qquad r \in (0, \infty).
\end{equation}

$\hat{h}^{(1)}$, $~v^{(1)}$, $~\hat{h}^{(2)}$, and
$~v^{(2)}~$ denote the solutions of two linearly
independent initial value problems for the equations
of motion  in the fields $~\hat{h}~$ and $~v$ (Jacobi
equation). The initial values are chosen such that they
are compatible with the symmetries of the background
solution.

A high order Runge-Kutta-Fehlberg method
\cite{StoBul90} has been employed for the numerical
integration of this system of differential equations.

\section{Results and Discussion}

First we discuss the results in the Prasad-Sommerfield
limit. The numerical analysis shows that the branch
$~BC~$ in {\small \bf Figure~1} is unstable in the sense of
linear perturbation theory, since the determinant has
a zero in the interior of the integration interval.

In {\small \bf Figure~2} the determinant (\ref{det}) is shown
for some values of the parameter \ad on the upper branch
branch. For $~\rum$, that is the value of $~r~$ where
the function $~\mu~$
has a minimum (see \cite{BreForMai92}), $~\det V(r)~$ attains
a local maximum. A second maximum at greater values of $~r$
is observed for some values of $~\alpha$. For \ad near
$~\a_\mx~$ the second maximum dominates,  whereas,
if  \ad is approximately $~\a_\mc$ the left peak is
well developed. For decreasing values of \ad
($\a \rightarrow \a_\mc$) the zero of
the determinant moves to the left.

It is convenient to rescale the determinant by
$~\mu(r) \mbox{e}^{-\a r}~$ in order to present
the functions for all \ad-values of the upper branch
in one plot (see {\small \bf Figure~3}). The pronounced
minimum in the left part of {\small \bf Figure~3} is due to
the rescaling of $~\det V(r)$. It suggests that
$~\lim_{\a \rightarrow \a_\mc}
\mu(\rum) \det V(\rum) = 0$, i.e. in the neighbourhood
of this point the determinant increases more slowly
than $~1/ \mu$.

For $~\a_\mx~$ the determinant has a zero at infinity in
agreement with a theorem in \cite{Str84} which
states that a change of the stability properties can
only happen at a point of bifurcation in the masses.
{}From this theorem it follows that the whole lower branch
in {\small \bf Figure~1} consists of stable
solutions, because of the stability of the Prasad-Sommerfield
solution in flat space ($\a = 0$).

The bifurcation point is determined using
Jacobi's criterion.
For increasing values
of \bd the method becomes more and more sensitive which means
that the error becomes smaller and smaller.

The determinant corresponding to solutions on the lower branch
is plotted in {\small \bf Figure~4}. As expected these functions
have no zeros. The curve with the steepest ascent belongs
to $~\a = 0$.

{\small \bf Figure~5} and {\small \bf 6} show the results for
some selected values of $~\b \neq 0$. One finds the same
qualitative behaviour as for $~\b = 0$.  In {\small \bf Figure~5}
the determinant is plotted for given \bd and the corresponding
maximal $~\a$-values. The lowest curve belongs to
$~\b = 0.05$. {\small \bf Figure~6} uses the same
$~\b$-values, but $~\a$-values on the upper branch.
Increasing \bd the zero of the determinant moves to
the left.

Some numerical results are collected in the following
table. $\a_\mc~$ is obtained by linear extrapolation:

\vspace*{1cm}
\begin{center}
\begin{tabular}{l|lll|l} \hline
$\b$ & $\a_\mx$ & $\mu_{\a_\mx}(\rum) $ & $M_{\rm max}$
& $\a_\mc$ \\ \hline \hline
0.00 & 1.403034 & $3.611 \cdot 10^{-2}$ &1.000219 & 1.3859  \\
0.05 & 1.397555 & $2.674 \cdot 10^{-2}$ &1.000138 & 1.3822  \\
0.10 & 1.389459 & $2.028 \cdot 10^{-2}$ &1.000076 & 1.3806  \\
0.15 & 1.379974 & $1.402 \cdot 10^{-2}$ &1.000038 & 1.3745  \\
0.20 & 1.369676 & $9.211 \cdot 10^{-3}$ &1.000017 & 1.3664  \\
0.25 & 1.358883 & $5.811 \cdot 10^{-3}$ &1.000007 & 1.3571  \\
\hline
\end{tabular}
\end{center}

\section*{Acknowledgement}\addcontentsline{toc}
{section}{Acknowledgement}

The author is indebted to P. Breitenlohner and D. Maison for
numerous discussions on the subject and to H. Ewen, P. Forg\'acs
and B. Schmidke for reading the manuscript.
She would also like to thank George Lavrelashvili.


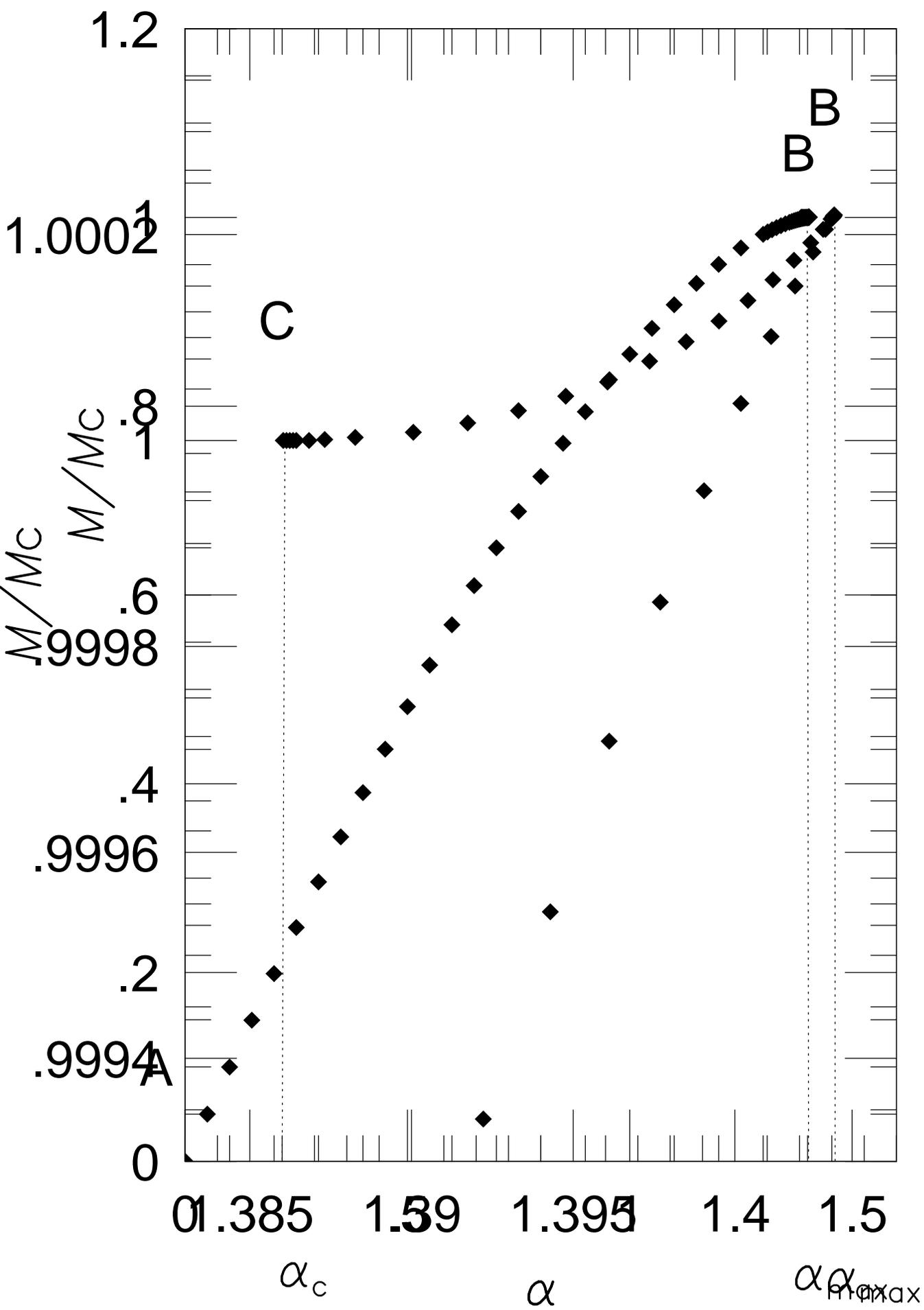
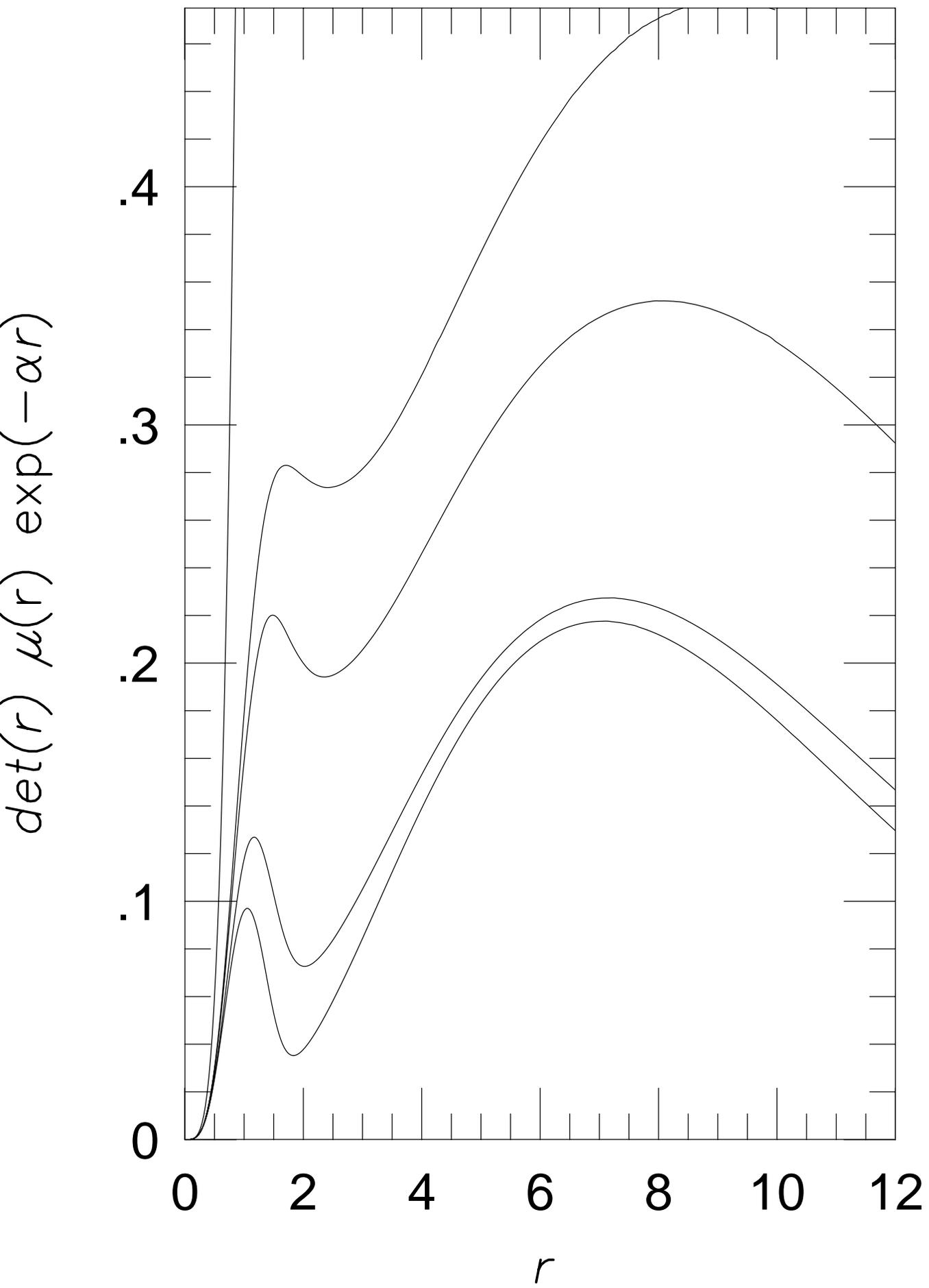
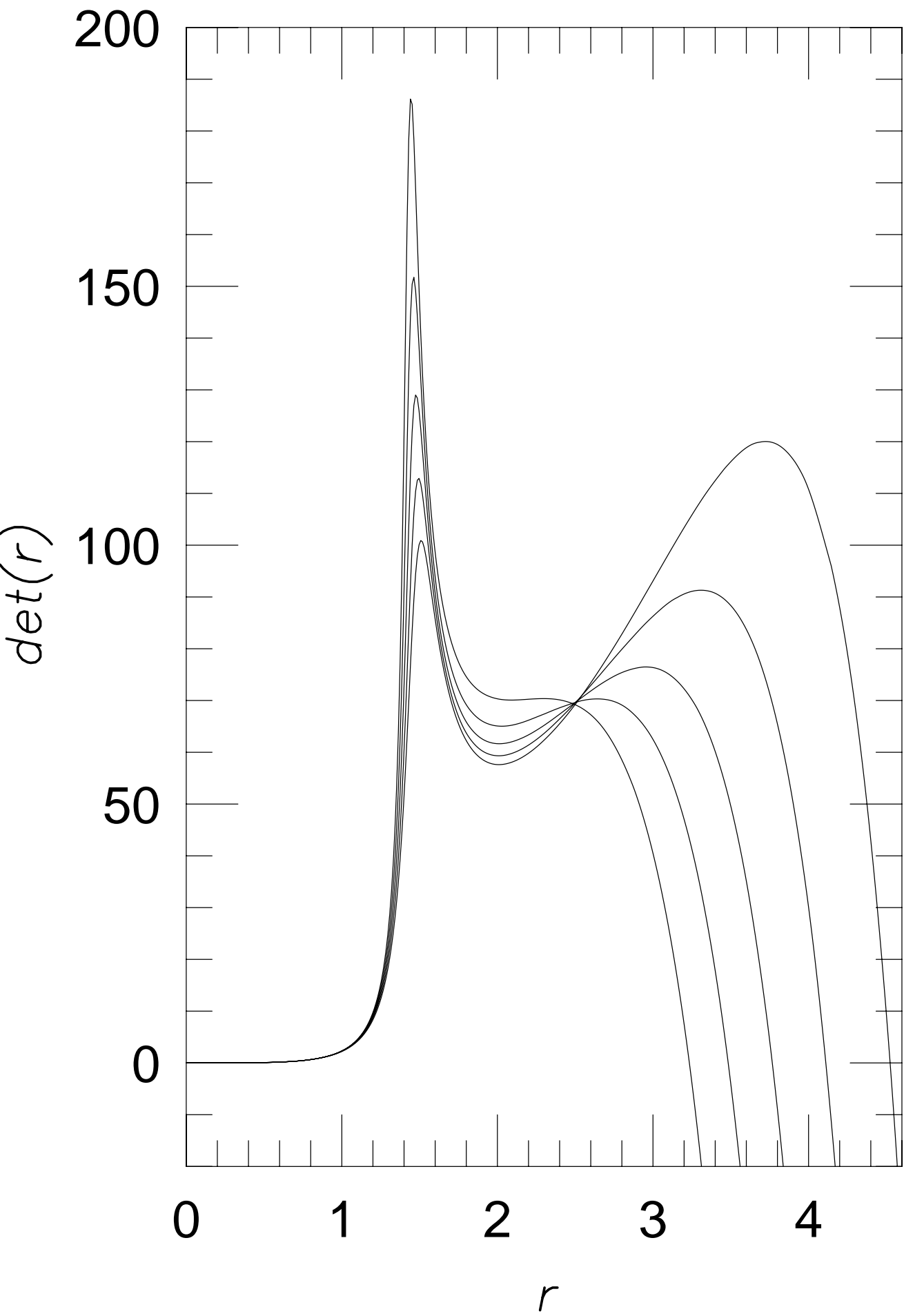
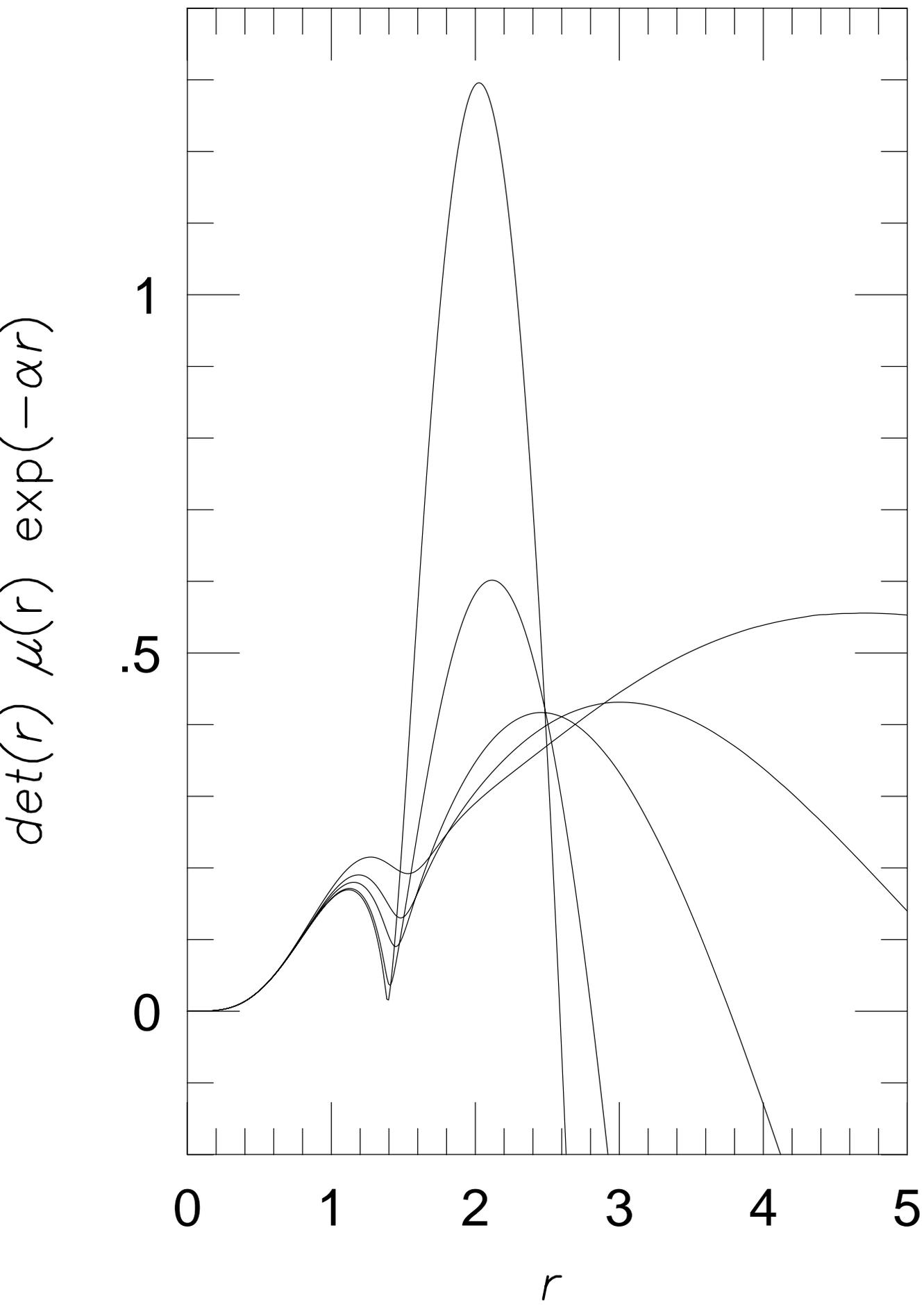
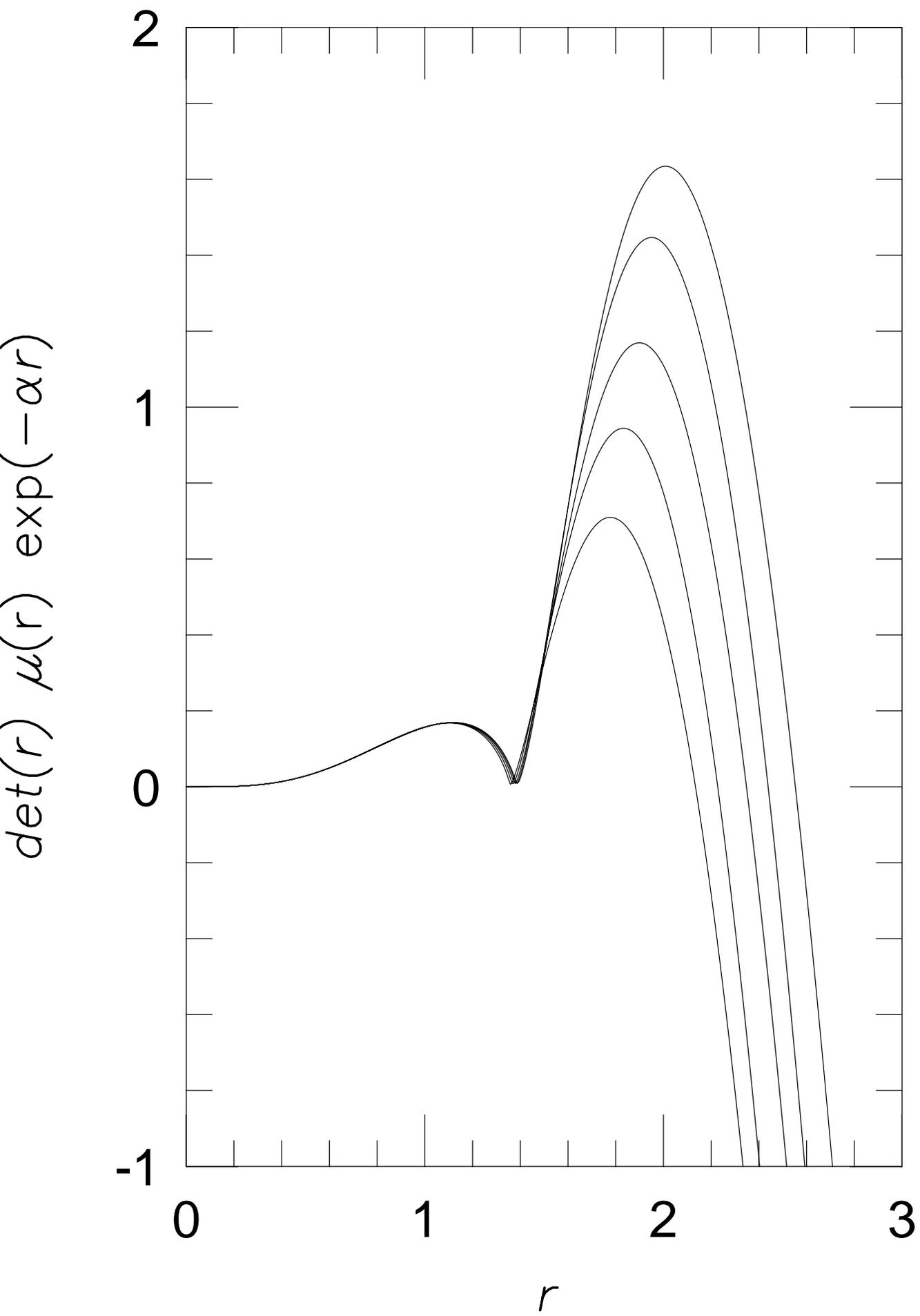
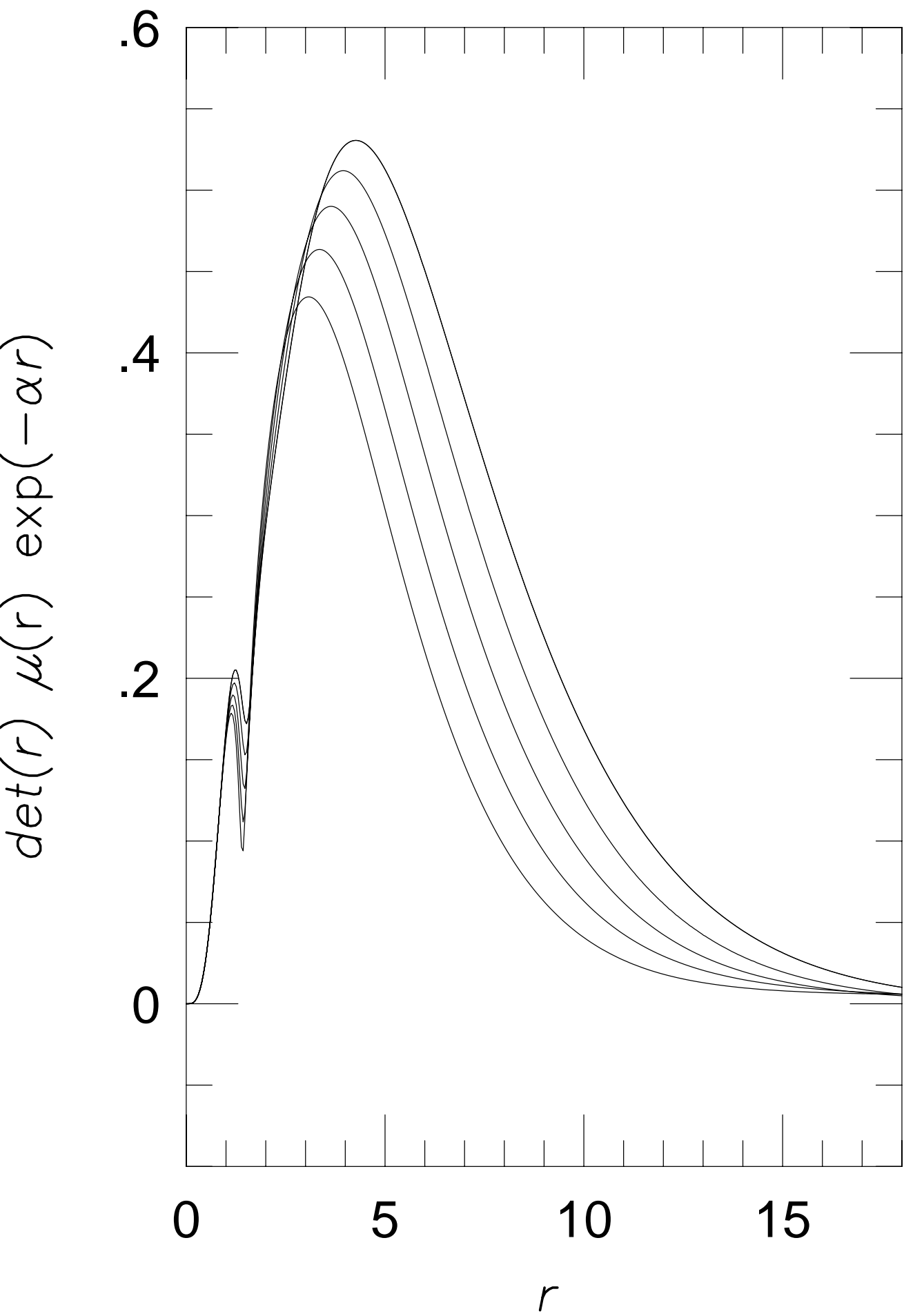
\end{document}